\documentclass[12pt]{article}
\usepackage{amssymb,graphicx,amsmath,amsfonts,graphics,xfrac,authblk}
\begin{document}

\title{Scalar Cosmological Perturbations}

\author[1]{Truman Tapia}

\author[2]{Clara Rojas}
\affil[1,2]{School of Physical Sciences and Nanotechnology, Yachay Tech University, 100119 Urcuqu\'i, Ecuador}
	
\maketitle

\begin{abstract}
In this paper we present the study of the scalar cosmological perturbations of a single field inflationary model up to first order in deviation.  The Christoffel  symbols and the tensorial quantities are calculated explicitly  in function of the cosmic time $t$. The Einstein equations are solved up-to first order in deviation and the scalar  perturbations equation is derived. 
\end{abstract}

\vspace{2pc}

\section{Introduction}
\begin{figure}[htbp]
\begin{center}
\includegraphics[scale=1]{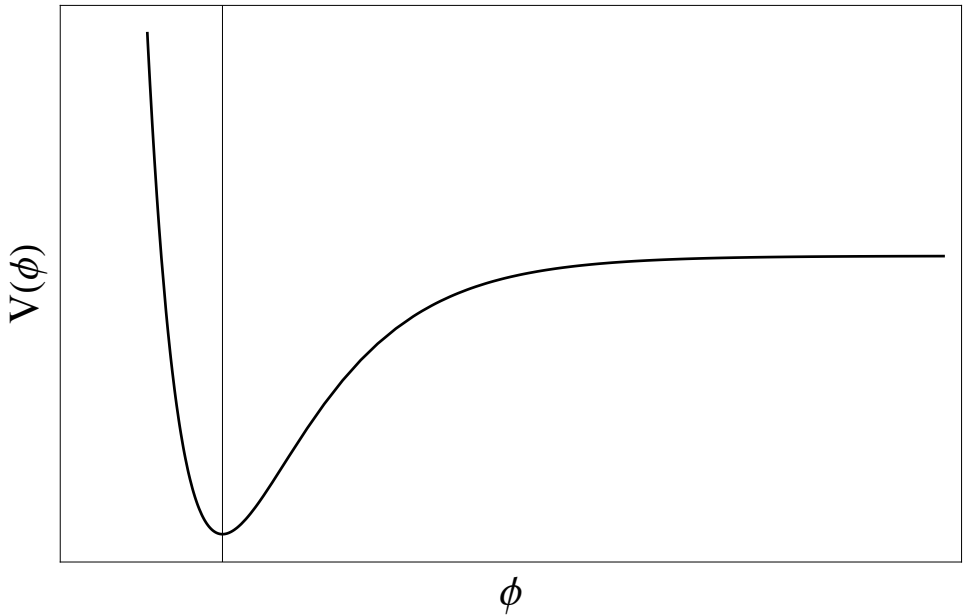}\\
\caption{The Starobinsky inflationary model.}
\label{infl_V}
\end{center}
\end{figure}

The inflationary theory was proposed in $1981$ by Alan Guth \cite{guth:1981} as a solution to the problems of the Big Bang theory.
The inflation is defined as a period of evolution of the Universe where the expansion was accelerated, so $\ddot{a}(t)>0$, being $a(t)$  the scale factor.
The fundamental characteristic of this theory is a period of expansion extremely fast in a very short period of time that happened when the Universe was extremely young \cite{liddle:2000}. The condition for inflation also can be written as $p<-\frac{\rho}{3}$, so to produces inflation we need matter with the property of having negative pressure. A matter that has this property is a scalar field $\phi$, called inflaton. The scalar field has associate a density of energy $\rho(t)$ and a potential $V(\phi)$, so when the potential  $V(\phi)$ dominates over the kinetic term  $\dot{\phi}^2$ we have inflation. This is possible if you have a sufficiently flat potential, as for example the Starobinsky potential. Figure \eqref{infl_V} shows the Starobinsky potential $V(\phi)= M^4 \left(1-e^{\sqrt{\sfrac{2}{3}}\phi}/M_{\textnormal{Pl}}\right)^2$ \cite{martin:2019},  which is the inflationary model supported by the currently observations \cite{akrami:2018} of the Planck satellite.

The observations of the CMB indicate that the universe was extremely homogeneous and isotropic in early ages. If the universe was so homogeneous? Where did galaxies form? It is assumed that there were small primordial perturbations that grew in amplitude due to gravitational instability to form the structures observed today.
Although,  at first stage,  the inflationary cosmology solved the fine-tuning problems of the Hot Big Bang model, its most important property  is that produces a spectrum of density perturbations or scalar perturbations which are responsible for galaxy formation and CMB anisotropies. So, it was realized that inflation can explain the large scale structure of the Universe \cite{guth:1982}. In fact, this theory points that quantum vacuum fluctuations are stretched by inflation, to then being a seed of perturbations, which will form the  large scale structure of the Universe through gravitational instability. Even more, most models of inflation predicts that the power spectrum of perturbations must be almost scale invariant. Measurements on the CMB radiation done by Planck mission have tested the success of this prediction, and so the importance of Inflation \cite{akrami:2018}.

Perturbations during inflation can be studied through the theory of cosmological perturbations; mathematically this is reduced to solve the Einstein equations up-to first order over a background, an expanding homogeneous universe describes for the FRW metric for a flat universe and characterized for a scale factor $a(t)$. One of the first works about the theory of cosmological perturbations was done by Bardeen \cite{bardeen:1980} then, it was studied extensively by Kodama and Sasaki \cite{kodama:1984} and, by Mukhanov \textit{et al.}  \cite{mukhanov:1992}.  Deruelle \textit{et al.} made a briefly revision on classical and quantum perturbations. Later some authors are discussed about the quantum origin of the cosmological perturbations \cite{kiefer:2009,sudarsky:2010,martin:2012}.
In the last years these theory  have been revised by several authors, up-to first order in deviation \cite{riotto:2002,sriramkumar:2009,christopherson:2010}, up-to second order in deviation \cite{acquaviva:2003,malik:2009,christopherson:2011} and, up-to third order in deviation \cite{christopherson:2009}.

The motivation of this article is to show explicitly the calculation of the Christoffel  symbols and the tensorial quantities in terms of the cosmic time $t$ for the perturbed metric. Although the linearization procedure is well-known from the literature on Cosmology, the explicit calculations are not shown, so this article serves as a  new source for students who are studying this topic for the first time.

In the following the Einstein summation convention is used. Greek indices run from 0 to 3, and Latin indexes run from 1 to 3. The signature of the metric $g_{\mu\nu}$ is given by $(-, +, +, +)$. Dot ($\dot{x}$) denotes differentiation with respect to the cosmic time $t$, and primes $(x^{'})$ differentiation with respect to the conformal time $\eta$. It is possible to go from time to conformal time using $dt=a \mathrm{d}\eta$. We  will only keep the terms up-to first order in approximation, i.e., linear in $\Phi$ and $\Psi$. 

In section 2 we discuss the background equations in inflation. In  section 3 we specify the line element for scalar perturbations, we calculated explicitly the Christoffel symbols and the tensorial quantities up-to first order in deviation. Section 4 is devoted to derive the equation for scalar perturbations. Finally, in section 5 we present our conclusions.

\section{Background equations of motion}

The line element of a homogeneous, isotropic and spatially flat universe is given by the Friedmann-Robertson-Walker metric, which is given by

\begin{equation}
\label{metric_FRW}
\mathrm{d} s^2= -\mathrm{d} t^2+a^2\delta_{ij}\mathrm{d} x^i\mathrm{d} x^j.
\end{equation}

From the FRW metric using the Einstein field equations we can obtain the background equations of motion, which are later used to study scalar perturbations. The starting point are the Friedmann equation for a flat universe without cosmological constant  Eq. \eqref{friedmann} and, the continuity  equation Eq. \eqref{continuity}

\begin{equation}
\label{friedmann}
\left(\dfrac{\dot{a}}{a}\right)^2=\dfrac{8\pi G}{3} \rho,
\end{equation}
	
\begin{equation}
\label{continuity}
\dot{\rho}=-3\dfrac{\dot{a}}{a}(\rho +P).
\end{equation}
$P$ and $\rho$ must be express in terms of a single scalar field (inflaton) and its potential through \cite{liddle:2000},

\begin{equation}
\label{rho}
\rho=\dfrac{1}{2}\dot{\varphi}_0^{2}+V(\varphi_0),
\end{equation}
	
\begin{equation}
\label{P}
P=\dfrac{1}{2}\dot{\varphi}_0^{2}-V(\varphi_0),
\end{equation}
where the potential $V(\varphi_0)$ specifies the model of inflation used. Replacing \eqref{rho} and \eqref{P} in equations \eqref{friedmann} and \eqref{continuity} results in the two background equations

\begin{equation}
\label{H}
H^2=l^2\left[\frac{1}{2}\dot{\varphi_{0}}^2+V(\varphi_{0})\right],
\end{equation}
and,
\begin{equation}
\label{varphi}
\ddot{\varphi_{0}}+3H\dot{\varphi_{0}}+V_{,\varphi_{0}}=0,
\end{equation}
where, $l^2=\sfrac{8\pi G}{3}$, $H=\sfrac{\dot{a}}{a}$, and $V_{,\varphi_{0}}$ represents differentiation of the potential with respect to $\varphi_{0}$.

\noindent Differentiating Eq. \eqref{H} with respect to time gives,
\begin{equation}
\label{2H}
2H\dot{H}=l^2(\dot{\varphi_{0}}\ddot{\varphi_{0}}+V_{,\varphi_{0}}\dot{\varphi_{0}}),
\end{equation}

\noindent multiplying  Eq. \eqref{varphi} by $\dot{\varphi_{0}}$ results in
\begin{equation}
\label{dotvarphi}
\dot{\varphi_{0}}\ddot{\varphi_{0}}+3H\dot{\varphi_{0}}^2+\dot{\varphi_{0}}V_{,\varphi_{0}}=0.
\end{equation}

Subtracting Eqs. \eqref{2H} and \eqref{dotvarphi} yields to

\begin{equation}
\dot{H}=-\frac{3}{2}l^2\dot{\varphi_{0}}^2,
\end{equation}
which in conformal time is

\begin{equation}
\label{H2}
    \mathcal{H}^2-\mathcal{H}'=\frac{3}{2}l^2\varphi_{0}'^2.
\end{equation}

\section{Scalar Perturbations up-to first order}
 

We are going to consider perturbations around the FRW universe under the longitudinal gauge, also known as conformal Newtonian gauge. The corresponding line element is given by,
	
\begin{equation}
\label{line_scalar}
\mathrm{d} s^2=-\left[1+2\Phi(t,\mathbf{x})\right] \mathrm{d} t^2+a^2\left[1-2\Psi(t,\mathbf{x})\right]\delta_{ij}\mathrm{d} x^i\mathrm{d} x^j,
\end{equation}
where  $\Phi(t,\mathbf{x})$ corresponds to the Newtonian potential and $\Psi(t,\mathbf{x})$ to the perturbation of the spatial curvature. Under this gauge $\Phi$ and $\Psi$ are gauge invariant variables \cite{mukhanov:1992}, and in absence of them equation \eqref{line_scalar}  is reduced to the FRW line element \eqref{metric_FRW}.   
	
We have to solve the linearized Einstein equation in order to find the equation of perturbations, which is given by 
	
\begin{equation}
\label{Einstein}
\delta G^\mu_\nu\equiv \delta R^\mu_\nu-\dfrac{1}{2} \delta^\mu_\nu \delta\mathcal{R}=8\pi G \delta T^\mu_\nu ,
\end{equation}
where $\delta  G{^{\mu}_{\nu}}$ is the perturbed Einstein tensor, $\delta  T{^{\mu}_{\nu}}$ is the perturbed energy-momentum tensor, $\delta R^\mu_\nu$ is the perturbed Ricci tensor, $g^\mu_\nu$ is the metric tensor, and $\delta \mathcal{R}$ is the perturbed Ricci scalar.

\subsection{The metric}
	
From Eq. \eqref{line_scalar} the metric considering perturbations is given by,
	
\begin{equation}
\label{metric_scalar}
g_{\mu\nu}=
\begin{pmatrix} 
-1-2\Phi(t,\mathbf{x}) & 0            & 0            & 0\\
0        & a^2[1-2\Psi(t,\mathbf{x})] & 0            & 0\\
0        & 0            & a^2[1-2\Psi(t,\mathbf{x})]& 0\\
0        & 0            & 0            & a^2[1-2\Psi(t,\mathbf{x})]
\end{pmatrix}.
\end{equation}

The $\{00\}$ covariant component of the perturbed metric \eqref{metric_scalar} is given by,
	
\begin{equation}
g_{00}= -1-2\Phi.
\end{equation}
	
\medskip
The $\{ii\}$ covariant components becomes,
	
\begin{align}
g_{ii}=a^2(1-2\Psi). 
\end{align}
	
\bigskip
As the  covariant metric  $g_{\mu\nu}$ is a diagonal matrix the contravariant metric  $g^{\mu\nu}$ is given by the inverse of each component.  The $\{00\}$ contravariant component of the perturbed metric \eqref{metric_scalar} is given by,
	
\medskip
\begin{equation}
g^{00}=-\dfrac{1}{1+2\Phi} \sim -1+2\Phi.
\end{equation}
	
The $\{ii\}$ contravariant components becomes,
	
\medskip
\begin{align}
g^{ii}=\dfrac{1}{a^2(1-2\Psi)}\sim \dfrac{1}{a^2}(1+2\Psi). 
\end{align}
	
\subsection{Christoffel Symbols}
	
The Christoffel symbols are given by \cite{dodelson:2020},
	
\begin{equation}
\Gamma^\alpha_{\mu\nu}=\dfrac{1}{2}g^{\alpha\beta}\left(g_{\beta\mu,\nu}+g_{\beta\nu,\mu}-g_{\mu\nu,\beta}\right).
\end{equation}

 \noindent We are going to calculate all the Christoffel Symbols for the perturbed metric \eqref{metric_scalar}.  In the following calculation  we remain the terms up-to first-order, so we discard  terms like $\Phi\Phi_{,0}$, $\Phi\Phi_{,i}$, being the time-time compontent:

\begin{eqnarray}
\nonumber
\label{G000}
\Gamma^0_{00}&=&\dfrac{1}{2} g^{00} g_{00,0},\\
\nonumber
&=&\dfrac{1}{2}\left(-1+2\Phi\right)\left(-2\Phi_{,0}\right),\\
&\sim& \Phi_{,0}.
\end{eqnarray}

The space-time components,

\bigskip
\begin{eqnarray}
\nonumber
\label{G00i}
\Gamma^0_{0i}&=&\Gamma^0_{i0},\\
\nonumber
&=&\dfrac{1}{2} g^{00} g_{00,i},\\
\nonumber
&=&\dfrac{1}{2}\left(-1+2\Phi\right)\left(-2\Phi_{,i}\right),\\
&\sim&\Phi_{,i}.
\end{eqnarray}
	
\bigskip
\begin{eqnarray}
\label{G0ij}
\nonumber
\Gamma^0_{ij}&=& \Gamma^0_{ji},\\
\nonumber
&=&\dfrac{1}{2} g^{00}\left(-g_{ij,0}\right),\\
\nonumber
&=&-\dfrac{1}{2}\delta_{ij}\left(-1+2\Phi\right)\left[a^{2}\left(1-2\Psi\right)\right]_{,0},\\
\nonumber
&\sim&\delta_{ij}a^2\left[H-2H\left(\Psi+\Phi\right)-\Psi_{,0}\right],
\end{eqnarray}
where $H=\sfrac{\dot{a}}{a}$.

\bigskip
\begin{eqnarray}
\label{Gi00}
\nonumber
\Gamma^i_{00} &=& \dfrac{1}{2} g^{ij} \left(-g_{00,j}\right),\\
\nonumber
&=&\dfrac{1}{2} \delta_{ij} a^{-2}\left(1+2\Psi\right)\left(2 \Phi_{,i}\right),\\
&\sim& \delta_{ij}  a^{-2} \Phi_{,i}.
\end{eqnarray}
	
\bigskip
\begin{eqnarray}
\nonumber
\Gamma^i_{j0} &=& \Gamma^i_{0j}\\
\nonumber
&=&\dfrac{1}{2} g^{ik} \left(g_{jk,0}\right),\\
\nonumber
&=&\dfrac{1}{2} \delta_{ij}  a^{-2} \left(1+2 \Psi\right) \left[ a^{2}\left(1-2\Psi\right)\right]_{,0},\\
&\sim&\delta_{ij}(H-\Psi_{,0}).
\end{eqnarray}
	
\bigskip
\begin{eqnarray}
\nonumber
\Gamma^i_{jk} &=& \dfrac{1}{2} g^{il} \left(g_{jl,k}+g_{kl,j}-g_{jk,l}\right),\\
\nonumber
&=&\dfrac{1}{2} a^{-2} \left(1+2\Psi\right) \left[\delta_{ij} a^2\left(1-2\Psi\right)_{,k}+\delta_{ik}a^2(1-2\Psi)_{,j}-\delta_{jk} a^2\left(1-2\Psi\right)_{,i}\right],\\
\nonumber
&=&\left(1+2\Psi\right)\left(-\delta_{ij}\Psi_{,k}-\delta_{ik}\Psi_{,j}+\delta_{jk}\Psi_{,i}\right),\\
&\sim& -\delta_{ij} \Psi_{,k}-\delta_{ik} \Psi_{,j}+\delta_{jk} \Psi_{,i}.
\end{eqnarray}
	
\subsection{The Ricci tensor}
	
The Ricci tensor is defined as \cite{dodelson:2020}:
	
\begin{equation}
R_{\beta \nu}=\Gamma^{\mu}_{\beta\nu,\mu}-\Gamma^{\mu}_{\beta\mu,\nu}+\Gamma^{\mu}_{\sigma\mu}\Gamma^{\sigma}_{\beta\nu}-\Gamma^{\mu}_{\sigma\nu}\Gamma^{\sigma}_{\beta\mu}.
\end{equation}
	
\begin{flushleft}
For the calculation of the components of the Ricci tensor is useful to consider
\end{flushleft}
\begin{equation}
\mathcal{H}=\frac{a'}{a};\indent \indent H=\frac{\mathcal{H}}{a};\indent \indent \frac{\ddot{a}}{a}=\frac{\mathcal{H}'}{a^2}.
\end{equation}
	
\medskip
\begin{flushleft}
The time-time component of the  perturbed Ricci tensor is
\end{flushleft}
	
\bigskip
\begin{eqnarray}
\nonumber
R_{00}&=&\Gamma^{\mu}_{00,\mu}-\Gamma^{\mu}_{0\mu,0}+\Gamma^{\mu}_{\sigma\mu}\Gamma^{\sigma}_{00}-\Gamma^{\mu}_{\sigma0}\Gamma^{\sigma}_{0\mu},\\
\nonumber
&=&\Gamma^{i}_{00,i}-\Gamma^{i}_{0i,0}+\Gamma^{i}_{0i}\Gamma^{0}_{00}-\Gamma^{i}_{00}\Gamma^{0}_{0i}+\Gamma^{i}_{ji}\Gamma^{j}_{00}-\Gamma^{i}_{j0}\Gamma^{j}_{0i},\\
\nonumber
&=&a^{-2}\Phi_{,ii}-3\left(\frac{\ddot{a}}{a}-H^2-\Psi_{,00}\right)+3H\Phi_{,0}-3\left(H^2-2H \Psi_{,0}\right).\\
\nonumber
\delta R_{00}&=&\frac{\Phi_{,ii}}{a^2}+3\Psi_{,00}+3H\left(\Phi_{,0}+2\Psi_{,0}\right),\\
\nonumber
&=&a^{-2}\left[\nabla^2\Phi+3\Psi''+3\mathcal{H}\left(\Phi'+\Psi'\right)\right].
\end{eqnarray}
	
	\bigskip
\begin{flushleft}
The time-space component of the  perturbed Ricci tensor is
\end{flushleft}

\bigskip
\begin{eqnarray}
\nonumber
R_{0i}&=&\Gamma^{\mu}_{0i,\mu}-\Gamma^{\mu}_{0\mu,i}+\Gamma^{\mu}_{\sigma\mu}\Gamma^{\sigma}_{0i}-\Gamma^{\mu}_{\sigma i}\Gamma^{\sigma}_{0\mu},\\
\nonumber
&=&\Gamma^{0}_{0i,0}-\Gamma^{0}_{00,i}+\Gamma^{j}_{0i,j}-\Gamma^{j}_{0j,i}+\Gamma^{j}_{0j}\Gamma^{0}_{0i}-\Gamma^{j}_{0i}\Gamma^{0}_{0j}+\Gamma^{0}_{k0}\Gamma^{k}_{0i}-\Gamma^{0}_{ki}\Gamma^{k}_{00}\\
\nonumber
&+&\Gamma^{j}_{kj}\Gamma^{k}_{0i}-\Gamma^{j}_{ki}\Gamma^{k}_{0j},\\
\nonumber
&=&\dot{\Phi}_{,i}-\dot{\Phi}_{,i}+\delta_{ij} \left(H-\dot{\Psi}\right)_{,i}-\delta_{ij} \left(H-\dot{\Psi}\right)_{,i}+\delta_{ij}\left(H-\dot{\Psi}\right) \Phi_{,i}\\
\nonumber
&-&\delta_{ij} \left(H-\dot{\Psi}\right)  \Phi_{,j}+\delta_{ik} \Phi_k \left(H-\dot{\Psi}\right)-\delta_{ki} \left[H-2H\left(\Phi+\Psi\right)-\dot{\Psi}\right] \Phi_{,k}\\
\nonumber
&+&\left(-\delta_{jk} \Psi_{,j}-\delta_{ij} \Psi_{,i}+\delta_{ij} \Psi_{,j}\right)\left(H-\dot{\Psi}\right)- \left(-\delta_{kk} \Psi_{,i}-\delta_{ki} \Psi_{,k}+\delta_{ki} \Psi_{,k}\right)\left(H-\dot{\Psi}\right)\\
\nonumber
\delta R_{0i}&=&2\left(\dot{\Psi}_{,i}+H\Phi_{,i}\right)\\
&=&\frac{2}{a}\left(\Psi'_{,i}+\mathcal{H}\Phi_{,i}\right).   
\end{eqnarray}
	
\bigskip
\begin{flushleft}
The space-space component of the  perturbed Ricci tensor is
\end{flushleft}

\bigskip
\begin{eqnarray}
\nonumber
R_{ij}&=&\Gamma^{\mu}_{ij,\mu}-\Gamma^{\mu}_{i\mu,j}+\Gamma^{\mu}_{\sigma\mu}\Gamma^{\sigma}_{ij}-\Gamma^{\mu}_{\sigma j}\Gamma^{\sigma}_{i\mu}\\
\nonumber
&=&\Gamma^{0}_{ij,0}+\Gamma^k_{ij,k}-\Gamma^0_{i0,j}-\Gamma^{k}_{ik,j}+\Gamma^0_{00}\Gamma^0_{ij}+\Gamma^k_{0k}\Gamma^0_{ij}-\Gamma^0_{kj}\Gamma^k_{i0}-\Gamma^{k}_{0j}\Gamma^0_{ik}\\
\nonumber
&=&\delta_{ij}a^2\left\{H^2-2 H^2(\Phi+ \Psi)-4H\dot{\Psi}+\frac{\ddot{a}}{a}\left[1-2(\Phi+\Psi)\right]-2H\dot{\Phi}-\ddot{\Psi}\right\}\\
\nonumber
&-&2\Psi_{,ij}+\delta_{ij}\Psi_{,kk}-\Phi_{,ij}+3\Psi_{,ij}+\delta_{ij}a^2H\dot{\Phi}+\delta_{ij}a^2\left[H^2-2H^2(\Phi+\Psi)-2H\dot{\Psi}\right]\\
\nonumber
&=&\delta_{ij}a^2\left\{a^{-2}\nabla^2\Psi+\frac{\ddot{a}}{a}\left[1-2(\Phi+\Psi)\right]-H\dot{\Phi}-\ddot{\Psi}+2H^2-4H^2(\Phi+\Psi)-6H\dot{\Psi}\right\}\\
\nonumber
&-&(\Phi-\Psi)_{,ij}\\
\nonumber
&=&\delta_{ij}a^2\left\{a^{-2}\nabla^2\Psi+\mathcal{H'}[1-2(\Phi+\Psi)]-\mathcal{H}\Phi'-\Psi''+\mathcal{H}\Psi'+2\mathcal{H}^2-4\mathcal{H}^2(\Phi+\Psi)-6\mathcal{H}\Psi'\right\}\\
\nonumber
&-&(\Phi-\Psi)_{,ij}\\
\nonumber
&=&\delta_{ij}\left\{\nabla^{2}\Psi-\Psi''+[1-2(\Phi+\Psi)](\mathcal{H}'+2\mathcal{H}^{2})-\mathcal{H}(\Phi'+5\Psi')\right\}-(\Phi-\Psi)_{,ij}.\\
\end{eqnarray}

	\subsection{The Ricci scalar}
	
	\begin{flushleft}
		The Ricci scalar is defined as \cite{dodelson:2020}
	\end{flushleft}
	
	\begin{equation}
	\mathcal{R}=g^{\mu\nu}R_{\mu\nu}.
	\end{equation}
	
	\begin{flushleft}
		The first-order perturbed Ricci scalar is
	\end{flushleft}
	
		\begin{eqnarray}
		\nonumber
		\mathcal{R}&=&g^{00}R_{00}+g^{ij}R_{ij}\\
		\nonumber
		&=&(-1+2\Phi)\left[-3\frac{{\mathcal{H}'}}{a^2}+\frac{\nabla^2\Phi}{a^2}+3\Psi''+3H\left(\Phi'+\Psi'\right)\right]+a^{-2}(1+2\Psi)\\
		\nonumber
		&\times&\left\{3[\nabla^{2}\Psi-\Psi''+\left[1-2(\Phi+\Psi)\right](\mathcal{H}'+2\mathcal{H}^{2})-\mathcal{H}(\Phi'+5\Psi')]-(\Phi-\Psi)_{,ii}\right\}\\
\nonumber
		&=&-\frac{2}{a^2}\left[3\Psi''+\nabla^2(\Phi-2\Psi)+3\mathcal{H}(\Phi'+3\Psi')-3(\mathcal{H}'+\mathcal{H}^2)+6\Phi(\mathcal{H}'+\mathcal{H}^2)\right].\\
		\end{eqnarray}
	
	\subsection{The Einstein tensor}

The Einstein  tensor is defined as \cite{dodelson:2020}:

\begin{equation}
G^\mu_\nu=R^\mu_\nu-\dfrac{1}{2} g^\mu_\nu \mathcal{R}.
\end{equation}

The time-time component of the Einstein tensor is:

		\begin{eqnarray}
		\nonumber
		G_{00}&=&R_{00}-\frac{1}{2}g_{00}\mathcal{R}\\
		\nonumber
		&=&\frac{1}{a^2}\left[-3\mathcal{H}'+\nabla^2\Phi+3\Psi''+3\mathcal{H}(\Phi'+\Psi')\right]-\frac{(1+2\Phi)}{a^{2}}\\
		\nonumber
		&\times&\left[3\Psi''+\nabla^2(\Phi-2\Psi)+3\mathcal{H}(\Phi'+3\Psi')-3(\mathcal{H}'+\mathcal{H}^2)+6\Phi(\mathcal{H}'+\mathcal{H}^2)\right]\\
		\nonumber
		&=&\frac{1}{a^2}\left[-6\mathcal{H}\Psi'+2\nabla^2\Psi+3\mathcal{H}^2\right].\\
		\end{eqnarray}
Then, transforming it to a mixed tensor and linearizing with respect to the background 

\vspace{-0.2cm}
 
		\begin{eqnarray}
		\nonumber\\
		\nonumber
		G_{0}^{0}&=&g^{\mu0}G_{\mu0}=g^{00}G_{00}\\
		\nonumber
		&=&(-1+2\Phi)a^{-2}\left[-6\mathcal{H}\Psi'+2\nabla^2\Psi+3\mathcal{H}^2\right]\\
		\nonumber
		&=&a^{-2}\left[-3\mathcal{H}^{2}+6\mathcal{H}\Psi^{'}-2\nabla^2\Psi+6\mathcal{H}^{2}\Phi\right]\\
		\delta G_{0}^{0}&=&a^{-2}\left[6\mathcal{H}\Psi'-2\nabla^2\Psi+6\mathcal{H}^2\Phi\right].
		\end{eqnarray}

The time-space component of the Einstein tensor is:

		\begin{eqnarray}
		\nonumber
		G_{0i}&=&R_{0i}-\frac{1}{2}g_{0i}\mathcal{R}\\
		\nonumber
		&=&R_{0i}\\
		\nonumber
		&=&2a^{-1}\left(\Psi'_{,i}+\mathcal{H}\Phi_{,i}\right)\\
		\nonumber
		&=&\delta R_{0i}.
		\end{eqnarray}

	In this case, all this tensor component is perturbation. Then, turning it to a mixed tensor,

		\begin{eqnarray}
		\nonumber
		\delta G_{i}^{0}&=&g^{\mu0}G_{\mu i}\\
		\nonumber
		&=&g^{00}G_{0i}\\
		\nonumber
		&=&\left(-1+2\Phi\right)2a^{-1}(\Psi_{,i}'+\mathcal{H}\Phi_{,i})\\
		&=&-2a^{-1}\left(\Psi'+\mathcal{H}\Phi\right)_{,i}.
		\end{eqnarray}

\bigskip
The space-space component of the Einstein tensor is:

\bigskip

		\begin{eqnarray}
		\nonumber
		G_{ij}&=&R_{ij}-\frac{1}{2}g_{ij}\mathcal{R}\\
		\nonumber
		&=&\delta_{ij}\left\{-\Psi''+\nabla^2\Psi+[1-2(\Phi+\Psi)](\mathcal{H}'+2\mathcal{H}^2)-\mathcal{H}(\Phi'+5\Psi')\right\}-(\Phi-\Psi)_{,ij}\\
		\nonumber
		&+&\delta_{ij}(1-2\Psi)\left[3\Psi''+\nabla^2(\Phi-2\Psi)+3\mathcal{H}(\Phi'+3\Psi')-3(\mathcal{H}'+\mathcal{H}^2)+6\Phi(\mathcal{H}'+\mathcal{H}^2)\right]\\
		\nonumber
		&=&\delta_{ij}\left[-(2\mathcal{H}'+\mathcal{H}^2)+2\Psi''+\nabla^2(\Phi-\Psi)+2(2\mathcal{H}'+\mathcal{H}^2)(\Psi+\Phi)+2\mathcal{H}(\Phi'+2\Psi')\right]\\
		\nonumber
		&+&(\Psi-\Phi)_{,ij}.
		\end{eqnarray}

\bigskip
Turning it to a mixed tensor and linearizing with respect to the background

\bigskip

\begin{eqnarray}
\nonumber
		G^{i}_{j}&=&g^{\mu i}G_{\mu j}\\
		\nonumber
		&=&a^{-2}(1+2\Psi)\{\delta_{ij}[-(2\mathcal{H}'+\mathcal{H}^2)+2\Psi''+\nabla^2(\Phi-\Psi)+2(2\mathcal{H}'+\mathcal{H}^2)(\Psi+\Phi)\\
		\nonumber
		&+&2\mathcal{H}(\Phi'+2\Psi')]+(\Psi-\Phi)_{,ij}\}\\
		\nonumber
		&=&a^{-2}\{\delta_{ij}[-(2\mathcal{H}'+\mathcal{H}^{2})+\nabla^2(\Phi-\Psi)+2\Psi''+2(2\mathcal{H}'+\mathcal{H}^{2})\Phi+2\mathcal{H}(\Phi'+2\Psi')]\\
		\nonumber
		&+&(\Psi-\Phi)_{,ij}\}\\
		\nonumber
		\delta
		\label{Gij}
		 G_{j}^{i}&=&a^{-2}\left\{\delta_{ij}[\nabla^2(\Phi-\Psi)+2\Psi''+2(2\mathcal{H}'+\mathcal{H}^{2})\Phi+2\mathcal{H}(\Phi'+2\Psi')]+(\Psi-\Phi)_{,ij}\right\}.\\
		\end{eqnarray}

	\subsection{The Energy-Momentum tensor}
	
	The energy-momentum tensor for a scalar field is given by \cite{mukhanov:1992}
	
	\begin{equation}
	T^{\mu}_{\nu}=\partial^{\mu}\varphi \partial_{\nu}\varphi-\left[\frac{1}{2}\partial^{\alpha}\varphi \partial_{\alpha}\varphi+V\right]\delta_{\nu}^{\mu}.
	\end{equation}
The form of the perturbed inflaton field is
\begin{equation}
    \varphi(t,\mathbf{x})=\varphi_{0}(t)+\delta\varphi(t,\mathbf{x}).
\end{equation}
Where, $\varphi_{0}(t)$ represents the background field, and $\delta\varphi(t,\mathbf{x})$ is the linear perturbation. And, the potential is expanded around $\varphi_{0}$ using Taylor expansion
\begin{equation}
    V(\varphi)=V(\varphi_{0})+(\varphi-\varphi_{0})V_{,\varphi}=V(\varphi_{0})+\delta V_{,\varphi}.
\end{equation}
	
The time-time component of the perturbed energy-momentum tensor is:

\bigskip
		\begin{eqnarray}
		\nonumber
		T^{0}_{0}&=&\partial^{0}\varphi \partial_{0}\varphi-\left[\frac{1}{2}\partial^{\alpha}\varphi \partial_{\alpha}\varphi+V(\varphi)\right]\delta_{0}^{0}\\
		\nonumber
		&=&g^{00}(\partial_{0}\varphi)^{2}-\left\{\frac{1}{2}\left[g^{00}(\partial_{0} \varphi)^{2}\right]+V(\varphi)\right\}\\
		\nonumber
		&=&(-1+2\Phi)(\dot{\varphi}_{0}+
		\dot{\delta\varphi})^2-\left[\frac{1}{2}(-1+2\Phi)(\dot{\varphi}_{0}+\dot{\delta\varphi})^2+V(\varphi)\right]\\
		\nonumber
		&=&\frac{1}{2}(-1+2\Phi)(\dot{\varphi}_{0}+\dot{\delta\varphi})^2-V(\varphi)\\
		\nonumber
		&=&-\frac{\dot{\varphi_{0}}^2}{2}-\dot{\varphi_{0}}\dot{\delta\varphi}+\Phi\dot{\varphi_{0}}^2-V(\varphi_{0})-V_{,\varphi}\delta\varphi9\\
		&=&a^{-2}\left[-\frac{\varphi{'}_{0}^2}{2}-a^2V(\varphi_{0})+\Phi\varphi{'}_{0}^2 -\varphi'_{0}\delta\varphi'-a^2\delta\varphi V_{,\varphi} \right].
		\end{eqnarray}

Linearizing with respect to the background

		\begin{eqnarray}
		\delta T^0_0&=&a^{-2}\left(\Phi\varphi'^2_{0}-\varphi'_{0}\delta\varphi'-a^2\delta\varphi V_{,\varphi}\right).
		\end{eqnarray}

\bigskip
The time-space component of the perturbed energy-momentum tensor is:

		\begin{eqnarray}
		\nonumber
		T_{i}^{0}&=&\partial^{0}\varphi \partial_{i}\varphi-\left[\frac{1}{2}\partial^{\alpha}\varphi \partial_{\alpha}\varphi+V\right]\delta^{i}_{0}\\
		\nonumber
		&=&g^{\mu0}\partial_{\mu}\varphi\partial_{i}\varphi\\
		\nonumber
		&=&g^{00}\partial_{0}\varphi\delta\varphi_{,i}\\
		\nonumber
		&=&(-1+2\Phi)\dot{\varphi}_{0}\delta\varphi_{,i}\\
		\nonumber
		&=&-a^{-1}\varphi_{0}'\delta\varphi_{,i}\\
		&=&\delta T_{i}^{0}.
	\end{eqnarray}

The space-space component of the perturbed energy-momentum tensor is:

\bigskip
		\begin{eqnarray}
		\nonumber
		T^{i}_{j}&=&\partial^{i}\varphi\partial_{j}\varphi-\left[\frac{1}{2}\partial^{\alpha}\varphi\partial_{\alpha}\varphi+V(\varphi)\right]\delta_{ij},\\
		\nonumber
		&=&-\left[\frac{1}{2}\left(g^{0\mu}\partial_{\mu}\varphi\partial_{0}\varphi\right)+V(\varphi)\right]\delta_{ij}\\
		\nonumber
		&=&-\left[\frac{1}{2}g^{00}\partial_{0}\varphi\partial_{0}\varphi+V(\varphi)\right]\delta_{ij},\\
		\nonumber
		&=&-\left[\frac{1}{2}(-1+2\Phi)(\dot{\varphi}_{0}+\dot{\delta\varphi})^{2}+V(\varphi)\right]\delta_{ij},\\
		\nonumber
		&=&\delta_{ij}\left[\frac{\dot{\varphi}_{0}^{2}}{2}-\Phi\dot{\varphi}_{0}^{2}+\dot{\varphi}_{0}\dot{\delta\varphi}-V(\varphi_{0})-\delta\varphi V(\varphi)_{,\varphi}\right],\\
		\label{Tij}
		&=&\delta_{ij}a^{-2}\left[\frac{(\varphi_{0}^{'})^{2}}{2}-a^{2}V(\varphi_{0})-\Phi(\varphi_{0}^{'})^{2}+\varphi_{0}^{'}\delta\varphi^{'}-a^{2}\delta\varphi V(\varphi)_{,\varphi}\right].
		\end{eqnarray}

	And, linearizing with respect to the background 

		\begin{eqnarray}
		\label{deltaTij}
	    \delta T^{i}_{j}&=&\frac{\delta_{ij}}{a^2}\left[-\Phi(\varphi_{0}^{'})^{2}+\varphi_{0}^{'}\delta\varphi^{'}-a^{2}\delta\varphi V(\varphi)_{,\varphi}\right].
	    \end{eqnarray}

	\section{Equation for scalar perturbations}
	
	We obtain the equations of motion for classical cosmological perturbations from the linearized Einstein equations. Comparing Eqs. \eqref{Gij} and \eqref{Tij} we get that $\Phi=\Psi$ if $i\neq j$. Then, the three resulting equations are
	
	\bigskip
	\begin{align}
	\label{Phi1}
	\nabla^2\Phi-3\mathcal{H}\Phi'-\left(\mathcal{H}'+2\mathcal{H}^2\right)\Phi&=\frac{3}{2}l^2\left(\varphi'_0\delta\varphi'+V_{,\varphi}\,a^2\delta\varphi\right),\\
	\label{Phi2}
	\Phi'+\mathcal{H}\Phi&=\frac{3}{2}l^2\varphi'_0\delta\varphi,\\
	\label{Phi3}
	\Phi''+3\mathcal{H}\Phi'+\left(\mathcal{H}'+2\mathcal{H}^2\right)\Phi&=\frac{3}{2}l^2\left(\varphi'_0\delta\varphi'-V_{,\varphi}\,a^2\delta\varphi\right),
	\end{align}
	
	\noindent after equating the $\{00\}$ components of the linearized Einstein equations, equation \eqref{H2} is used to get Eq. \eqref{deltaTij}. Equation \eqref{varphi} in conformal time can be written as
	
	\begin{equation}
	\label{deltaV}
	\frac{V_{,\varphi_{0}}}{\varphi'_{0}}=-\frac{1}{a^2}\left[\frac{\varphi''_{0}}{\varphi'_{0}}+2\mathcal{H}\right].
	\end{equation}
	
	Subtracting Eq. \eqref{Phi1} from Eq. \eqref{Phi3}, replacing $\delta\varphi$ from Eq. \eqref{Phi2}, and using Eq. \eqref{deltaV}; we get
	
	\begin{equation}
	    \Phi''-\nabla^2\Phi+\Phi\left[2\mathcal{H}-2\frac{\varphi_{0}''}{\varphi_{0}'}\right]+\Phi\left[2\mathcal{H}'-\frac{2\mathcal{H}\varphi''_{0}}{\varphi'_{0}}\right]=0.
	\end{equation}
	
	\noindent which can be express as
	
	\begin{equation}
	\label{Phi_final}
	\Phi''-\nabla^2\Phi+2\left(a/\phi'_0\right)'\left(a/\varphi_0'\right)^{-1}\Phi'+2\varphi'_0\left(\mathcal{H}/\varphi_0'\right)'\Phi=0.
	\end{equation}
	or,
	
	\begin{equation}
	\label{classical_eq}
	u''-\nabla^2u -z\left(\frac{1}{z}\right)''u=0,
	\end{equation}
	
	\noindent defining $u=\left(\sfrac{a}{\varphi'_0}\right)\Phi$ and $z=\sfrac{(a\varphi_0')}{\mathcal{H}}$.
	
	Equation \eqref{classical_eq} describes the evolution of perturbations classically, and can be used to study post-inflationary evolution after horizon crossing. To study perturbations inside horizon, it is necessary to quantize the perturbations, which is valid if we start from an action formalism \cite{mukhanov:1992}. But, one could also start from \eqref{classical_eq} and arrive to the equation describing perturbations inside horizon as Deruelle \cite{deruelle:1992} shows. 
	
	Thus, following Deruelle prescriptions consider the gauge invariant variable (which arises from Mukhanov derivation)
	
	\begin{equation}
	    v=a\left[\delta\varphi+\frac{\varphi_{0}\Phi}{\mathcal{H}}\right],
	\end{equation}
and its relations with $u$ given by
	
	\begin{equation}
	\label{eq_v}
	v=\frac{2}{3l^2}\left(u'+\frac{z^{'}}{z}u\right),
	\end{equation}
differentiating \eqref{eq_v} with respect to conformal time and using \eqref{classical_eq} we find that,
	
	\begin{equation}
	\label{nabla_u}
	\nabla^{2}u=\frac{3}{2}l^2z\left(\frac{v}{z}\right)'.
	\end{equation}

Taking $\nabla^{2}$ of \eqref{classical_eq},
	$$\nabla^{2}(u)''-\nabla^{2}(\nabla^{2}u)-z(1/z)''\nabla^{2}u.$$
then substituting \eqref{nabla_u} here
	
	$$\left[\left(v'-\frac{z'}{z}v\right)\right]''-\nabla^{2}\left[z\left(\frac{v}{z}\right)'\right]-\left(\frac{2z'}{z^{2}}-\frac{z''}{z}\right)\left(v'-\frac{z'}{z}v\right)=0,$$
which after solving we find that,
	
$$
z\left(\frac{v''-\nabla^{2}v-\frac{z''}{z}v}{z}\right)'=0.
$$
	
	\noindent Thus,

\begin{equation}
v''-\nabla^{2}v-\frac{z''}{z}v=z\times \textnormal{cons}
\end{equation}

\noindent but $\textnormal{cons}\neq0$ do not describe small perturbations, remaining

\begin{equation}
    v''-\nabla^{2}v-\frac{z''}{z}v=0,
\end{equation}
in the Fourier space it is obtained that the mode  $v_k$ with wave number $k$ satisfies the differential equation
	
	\begin{equation}
	\label{ddotukS}
	v''_k+\left(k^2-\frac{z''}{z}\right) v_k=0,
	\end{equation}
	with the asymptotic condition, coming from quantization, that the modes are initially plane waves for scales of length less than the Hubble horizon
	
	\begin{equation}
	\label{uk_boundary}
	\lim_{\eta\rightarrow-\infty} v_k\rightarrow \frac{1}{\sqrt{2k}} e^{-ik\eta}.
	\end{equation}
	
	
	
	
	
 	

	\section{Conclusions}
	\label{conclusions}
	 In this work we have done the study of scalar cosmological perturbations. We have explicitly made the calculations of the Christoffel symbols and the tensorial quantities in function of the cosmic time $t$. We have  solved the linearized Einstein equations and the scalar cosmological perturbation equation has been derived.

\section{References}
\bibliography{cp}
	

\end{document}